# Evaluating Patient Readmission Risk: A Predictive Analytics Approach

[1,2,3]Avishek Choudhury and [4]Dr. Christopher M. Greene

[1]*Applied Data Science, Syracuse University, New York, USA*
[2]*Process Improvement, UnityPoint Health, Iowa, USA*
[3]*Systems Engineering, Stevens Institute of Technology, New Jersey, USA*
[4]*Systems Science and Industrial Engineering, Binghamton University, NY, USA*

**Abstract:** With the emergence of the Hospital Readmission Reduction Program of the Center for Medicare and Medicaid Services on October 1, 2012, forecasting unplanned patient readmission risk became crucial to the healthcare domain. There are tangible works in the literature emphasizing on developing readmission risk prediction models; However, the models are not accurate enough to be deployed in an actual clinical setting. Our study considers patient readmission risk as the objective for optimization and develops a useful risk prediction model to address unplanned readmissions. Furthermore, Genetic Algorithm and Greedy Ensemble is used to optimize the developed model constraints.

## Introduction

It is a fact that the federal budget of the United States is concerned by the burgeoning healthcare expenses (Shipeng Yua, 2015). One of the main factors contributing to the healthcare cost is the avoidable patient readmission. Unplanned patient readmission has been a significant measure of care quality. However, the Affordable Care Act of 2010 introduced the Readmission Reduction Program which became effective on October 1, 2012. According to the School of Public Health, Veterans Administration can save $2,140 per patient by managing patients prone to readmission (Kathleen Carey, 2016).

Moreover, studies have shown that 15 to 25 percent of discharged patients are readmitted in less than 30 days. According to the Agency for Healthcare Research and Quality, about 1.8 million patients were readmitted (Anika and Hines, 2014). Fierce Healthcare reported that in 2011, hospitals spend $41.3 billion to treat unplanned readmitted patients (Shinkman, 2014).

A study published by Harvard Business Review stated that prioritized and effective communication with the patient and complying to evidence-based care standards could check patient readmission rate by 5 percent (Claire Senot, 2015). However, fostering desired communication within a hospital is arduous due to the complexity of the system.

Our study focuses on predicting patient readmission. Individuals with a high risk of readmission can be provided with alternative preventive measures such as intensive post-discharge care or home care (Davood Golmohammadi, 2015).

We define patient readmission as the readmission caused due to poor discharge planning resulting in reoccurrence of the treated disease and worsening health condition. When an individual requires readmission within 90 days' post-discharge for the same cause for which she or he was admitted to a hospital in the very first place is termed as the patient readmission. The reason behind considering readmission within 90 days is since the patients during the first three-month post-discharge are susceptible to the diseases and have suicidal behavior among individuals who have a mental disorder (Appleby, 2013).

## Alarming Hospital Discharge Concerns

This section classifies the three most crucial poor patient discharge issues that encourages patient readmission.

### *Early Patient Discharge*

The fundamental decision healthcare providers need to take is whether an individual has recovered enough to leave the hospital independently. Poor decision making at this instance hinders patient safety, resulting in emergency readmission or sometimes death. "A man died after a hospital failed to treat sepsis" and discharge the patient before time (Ombudsman, 2003). According to Homeless Link, more than 70% of underprivileged people were discharged without any housing and addressing underlying health conditions (IHSMHL, 2012).

*Poor Patient Assessment and Consulting Prior Discharge*

Often patients physically fit enough are not mentally capable of coping at home. These patients after discharge often fail to continue medications and lose mental health which in turn enhances the plausibility of readmission. Such conditions are common among older adults who are not capable of independently maintaining their health either due to cognitive or financial constraints. According to King's Fund, "being discharged without proper support is an invitation to relapse, worsening of the condition and readmission" (Maguire, 2015).

*Absence of Home Care Plans*

Insufficient communication and coordination between hospitals and community healthcare providers is another concern that needs attention. Due to insufficient domestic healthcare facilities, discharged patients with health care requirements are left alone at home which leaves the patient susceptible to health deterioration and emergency readmission. During 2002 and 2012, 3,225 suicides were recorded by The National Confidential Inquiry into Suicide and Homicide by People with Mental Illness, 2014. To minimize such occurrences, NHS recommends hospitals to follow up with their discharged patients within 7-2002 1day post discharge and ensure availability of crisis support (Assessment, 2013).

## Problem Statement

There exist several possible causes responsible for unplanned patient readmission. However, our study does not focus on identifying the responsible cause, but it provides with an efficient prediction model that can be deployed to a clinical scenario and help healthcare units to be prepared for the unavoidable readmissions and provide alternative care to preventable readmissions. The proposed model provides healthcare providers with a decision support system to identify individuals prone to readmission and thus minimize early discharge and ensure follow up with the discharge patients.

## Systematic Literature Review

*Design*

The study conducts a systematic review of methods and models used in predicting unplanned patient readmission to meet our research motive (Dixon Woods, 2005). The literature search process comprised of the following three steps: (a) Systematic literature search using electronic database search, 'snowballing' (Greenhalgh, 2005), (b) identify relevant papers based on their title and (c) article selection based on their abstract.

*Search Strategy and Inclusion Criteria*

The literature survey was limited to the following database: ACM Digital Library, ASME Digital Collection, BIOSIS Citation Index, CINDAS Microelectronics Packaging Material Database, Cite-Seer, Computer Database, Emerald Library, Energy and Power Source, Engineering Village, IEEE Xplore, MEDLINE, OSA Publishing, PubMed, Safari Books Online, ScienceDirect, Sci-Finder, SPIE Digital Library and Springer. The search conducted had no constraint of time zone and the following material type was considered: Articles, Newspaper Articles, Dissertations, Conference Proceedings, Statistics Data Sets, Technical Reports and Websites. The search keywords used were "patient readmission," "readmission" risk, "readmission survey," "readmission prediction" and "prediction models." All the papers that contained any of these words anywhere in the article were selected. Then based on their title, 104 papers where shortlisted. Finally, after reading the abstract 33 peers reviewed articles were finalized as the reference for this study.

*Findings From the Literature Review*

Miller *et al*. (1984) used multiple regression to develop a five-year prediction model for patient readmission. This paper was an indication that multiple regression for predictor variable analysis was a viable option. Hodgson *et al*. (2001) estimated readmission rates for all psychiatric admissions in North Staffordshire. Survival analysis was used to find the ones that predicted readmission. It used the Survival Analysis Log-Rank Test and Cox Regression for this purpose. Betihavas *et al*. (2012) extended the scope of the study done. It included non-clinical factors as well such as social instability. It also called out the need for predictive analysis and the lack of such tools for clinicians to use in risk assessment of readmission. Allison (2012) studied variables that could potentially predict readmission chances for patients previously admitted in pulmonary rehabilitation. Apart from eliminating excessive pain and unnecessary illness, it sought to reduce health care cost, in general, using discriminant and predictive analysis. Zheng *et al*. (2015) used metaheuristic and data mining approaches. A neural networks algorithm and SVM classifiers were used. These models could perform risk measure with higher sensitivity and F measure.

Bakal *et al*. (2014) generated a prediction as well as risk evaluation model for the rehospitalization of Heart Failure (HF) patients. It was found using five years of follow up half the patients had returned for rehospitalization. It concluded that trying to elongate the gap between hospitalizations should be an essential goal for evaluating the quality of treatment. Ajorlou *et al*. (2014) proposed a risk prediction model based on hierarchical nonlinear mixed effect to recognize patients with high likelihood of discharging, non-compliances to decrease Medicare costs and improve quality of care provided by hospitals. It applied stepwise variable selection in the mixed-effect framework and extended the (typical) random frailty model for Weibull hazard function with incorporated patient factors. Wang *et al*. (2014) validated the use of the LACE index when studying readmission risk of patients with CHF. Bayati *et al*. (2014) evaluated the cost-effectiveness and efficiency of the methodology that combines prediction and decision making. Machine learning classifiers were used with the patient data to perform the cost analysis. Inouye *et al*. (2015) was one of the few studies that used patient self-reports as means for risk assessment of readmission. An automated multi-call follows up system was implemented.

Amarasingham *et al*. (2015) in the same year focused on Electronic Medical Record (EMR) models to access readmission. Kang *et al*. (2015) used Retrospective Analysis and multivariate analysis to determine readmission risk factors. Futoma *et al*. (2015) compared several predictive models for predicting early readmissions. Deep learning is used to analyze the five conditions that CMS uses to penalize hospitals. It used Logistic Regression, Penalized Logistic Regression, Random Forests and Support Vector Machines and Neural Network deep learning methods. A framework for assessing patient readmission risk was developed. It found random forests, penalized regressions and deep neural networks to be the best predictors. Shams *et al*. (2015) developed a new metric for evaluating possible avoidable readmission. A tree-based classification method was proposed that factored in the previous history of the patient's readmissions and the various risk factors that were identified by the researchers of the paper. Pack *et al*. (2016) focused on readmission prediction for patients with heart valve surgery specifically. It used a generalized predictive equation for predicting readmission.

Turgeman and May (2016) developed a predictive model for hospital readmissions using a boosted C5.0 tree and Support Vector Machine as base and secondary classifiers respectively. It tried to balance the readmission classification problem. Lewis *et al*. (2016), compared the accuracy of two different risk prediction models, The Hospital Readmission Reduction Program (HRRP) and the Risk-Standardized Readmission Rate (RSRR) models. (Tong *et al*., 2016) compared several existing models on an all-cause non-elective basis. LACE, LASSO logistic, AdaBoost, STEPWISE logistic are compared with varying sample sizes. Golmohammadi used neural networks, classification and regression models and chi-square automatic interaction detection for analysis (Golmohammadi and Radnia, 2016). All models had an overall accuracy of over 80%.

The latter two gave the user the ability to select misclassification costs additionally. C5.0 was used to find any recurring patterns using patient history. Low *et al*. (2016) also compared the results with LACE index. After retrospective cohort analysis, like Kang *et al*. (2015), it grouped the predictors into categories. Wang *et al*. (2016), aimed to find the accuracy of Severity of Illness (SOI) and Risk of Mortality (ROM) individually, in predicting readmissions. Similar to Hogarth's work (Mahajan *et al*., 2016; 2017) created a regularized logistic regression model for risk prediction on a thirty-day basis. This was yet another study in the heart related patient readmission domain and was limited to risk prediction and comparison of risk prediction models specifically (Mahajan *et al*., 2016). Kroeger *et al*. (2018) determines whether Pediatric Early Warning Score before transfer may serve as a predictor of unplanned readmission to the cardiac intensive care unit. Jiang *et al*. (2018) utilized feature selection algorithms and machine learning models to develop a risk prediction system that is dynamic and accurate.

Several studies have implemented diverse modeling methods to determine the factors that influences hospital patient readmission rate (Betihavas *et al*., 2012; Davison *et al*., 2016; Golas *et al*., 2018; Hebert *et al*., 2014; Lum *et al*., 2012; SHAMEER *et al*., 2017; Shams *et al*., 2015; Wasfy *et al*., 2013; Yu *et al*., 2015).

## Methodology

Our study does not involve the participation of any patient. All analysis is based on anonymized data and ensures confidentiality. The dataset consists of 55 attributes and a sample size of 100,000 instances and represents 10 years of data collected from 130 US hospitals (Avishek, 2018; Strack *et al*., 2014). Table 1 below shows the data distribution. The original database contains curtailed, superfluous and noisy information as expected in most of the real-world data (Strack *et al*., 2014). There were some attributes that could not be treated directly since they had a high percentage of missing values. These features were "weight" (97% values missing), "payer code" (40%) and "medical specialty" (47%). "Weight" attribute was too sparse to be considered and was not included in further analysis. "Payer code" was neglected since it had a high percentage of missing values and it was not considered relevant to the outcome. "Medical specialty" attribute was accounted for analysis, adding the value "missing" in order to account for missing values. Large percentage of missing values of the "weight" attribute can be explained by the fact that prior to the HITECH legislation of the American Reinvestment and Recovery Act in 2009 hospitals and clinics were not required to capture it in a structured format (Strack *et al*., 2014).

**Table 1:** Data description

| Predictors | Number of encounters | % of population | Readmitted Number of encounters | % in group |
|---|---|---|---|---|
| **HbA1c** | | | | |
| No test was performed | 57,080 | 81.6% | 5,324 | 9.4% |
| Result was high and the diabetic medication was changed | 4,071 | 5.8% | 361 | 8.9% |
| Result was high, but the diabetic medication was not changed | 2,196 | 3.1% | 166 | 7.6% |
| Normal result of the test | 6,637 | 9.5% | 590 | 8.9% |
| **Gender** | | | | |
| Female | 37,234 | 53.2% | 3,462 | 9.3% |
| Male | 32,750 | 46.8% | 2,997 | 9.2% |
| **Discharge disposition** | | | | |
| Home | 44,339 | 63.4% | 3,184 | 7.2% |
| Otherwise | 25,645 | 36.6% | 3,275 | 12.8% |
| **Admission source** | | | | |
| Emergency room | 37,277 | 53.3% | 3,563 | 9.6% |
| Referrals | 22,800 | 32.6% | 2,032 | 8.9% |
| Otherwise | 9,907 | 14.2% | 846 | 8.5% |
| **Specialty of the admitting physician** | | | | |
| Internal medicine | 10,642 | 15.2% | 1,044 | 9.8% |
| Cardiology | 4,213 | 6.0% | 309 | 7.3% |
| Surgery | 3,541 | 5.1% | 284 | 8.0% |
| General practice | 4,984 | 7.1% | 492 | 9.9% |
| Missing values | 33,641 | 48.1% | 3,237 | 9.6% |
| Other | 12,963 | 18.5% | 1,093 | 8.4% |
| **Primary diagnosis** | | | | |
| Circulatory system | 21,411 | 18.5% | 1,093 | 8.4% |
| Diabetes | 5,747 | 8.2% | 529 | 9.2% |
| Respiratory system | 9,490 | 13.6% | 710 | 7.5% |
| Digestive system | 6,485 | 9.3% | 532 | 8.2% |
| Injury and poisoning | 4,697 | 6.7% | 524 | 11.2% |
| Musculoskeletal and connective tissue problem | 4,076 | 5.8% | 354 | 8.7% |
| Genitourinary system disease | 3,435 | 4.9% | 313 | 9.1% |
| Neoplasm | 2,536 | 3.6% | 239 | 9.4% |
| Other | 12,107 | 17.3% | 1,129 | 9.3% |
| **Race** | | | | |
| African American | 12,626 | 18.0% | 1,116 | 8.8% |
| Caucasian | 52,300 | 74.7% | 4,943 | 9.5% |
| Other | 3,138 | 4.5% | 256 | 8.2% |
| Missing | 1,920 | 2.7% | 144 | 7.5% |
| **Age** | | | | |
| Less than or equal to 30yrs. | 1,808 | 2.6% | 112 | 6.2% |
| 30-60 yrs. | 21,871 | 31.3% | 1,614 | 7.4% |
| Older than 60yrs. | 46,305 | 66.2% | 4,733 | 10.2% |
| Age | mean | median | 1$^{st}$ Qu | 2$^{nd}$ Qu |
| Age in years | 64.9 | 67 | 55 | 77 |
| **Time in hospital** | | | | |
| Days between admission and discharge | 4.3 | 3 | 2 | 6 |

The primary dataset contained numerous inpatient visits for some patients and the observations could not be considered as statistically independent, an assumption of the logistic regression model. We thus used only one encounter per patient; in particular, we measured only the first encounter for each patient as the primary admission and determined whether or not they were readmitted within 90 days.

Furthermore, we detached all encounters that resulted in either discharge to a hospice or patient death. After filtering out the data, we were left with 69,984 encounters that constituted the final dataset for analysis.

The methodology employed in this study can be broadly categorized into the following sections: Data preprocessing, implementation of predictive models and model optimization.

*Data Preprocessing*

Data preprocessing included three steps: Feature selection, handling outliers, data balancing and data partitioning.

*Feature Selection*

Large datasets hinder the speed of algorithms and even deteriorate classification accuracy (Kohavi and John, 1997). The concern raised due to data size is termed as the minimal-optimal problem (Nilsson, 2007). Our study employs Boruta algorithm and stepwise regression to determine the best features within the dataset.

Boruta algorithm is a wrapper developed on random forest classification algorithm (Liaw and Wiener, 2002). In this algorithm, the relevance of any attribute is retrieved as the loss of classification accuracy caused due to permutation of attribute values among objects. It calculates the shuffled correlations between the response and the attributes. It also computes the Z-score to determine attributes' relevance by dividing the mean accuracy loss by its standard deviation. In addition to Boruta, stepwise regression was also implemented.

Stepwise regression is designed as an automatic computational procedure in which the performance of the regression increases with increase in the input variable (Barnett *et al.*, 1975; Campolongo *et al.*, 2000). Stepwise regression is a different version of the forward selection in which after every step a variable is added, all selected attributes in the model are analyzed to determine any loss in relevance. If an irrelevant variable is found, it is blocked from the model. Stepwise regression mandates two significance levels: One for adding attributes and one for eliminating attributes. The cutoff plausibility for adding an attribute must be less than the cutoff probability for eliminating attributes to avoid an infinite loop trap (Mengchao Wang, 2013).

*Missing Value Imputation*

Median Absolute Deviation Method was used to address missing values (Avishek, 2018; Leys, 2013). This method helps to avoid any outliers within the dataset. The same concern can also be addressed by scaling and normalizing the dataset between 0 and 1. The Equations (1) used in this study to perform data normalization are given below:

$$Normalized(N_i) = (N_i - E_{min}) / (E_{max} - E_{min}) \quad (1)$$

Where:
$E_{min}$ = The minimum value for variable $E$
$E_{max}$ = The maximum value for variable $E$
*Note: If $E_{max}$ equals $E_{min}$ then $N_i$ equals 0.5

*Handling Outliers*

Median Absolute Deviation Method was used to address outliers (Avishek and Greene C., 2018) and (Leys, 2013). All missing values were replaced by the column mean.

*Data Balancing*

The entire data was randomly shuffled. Oversampling, undersampling and rose sampling methods were performed to balance the response variable (Al-Wesabi *et al.*, 2018). Figure 1 shows the comparison of different data sampling methods.

*Data Partitioning*

The balanced data was partitioned into training (70%) and testing (30%). All predictive models were fitted on the training data and the testing accuracy was measured for model evaluation.

*Implementation of Prediction Models*

Algorithms such as random forest, support vector machine, Recursive Partitioning and Regression Tree, Gradient Boosting Method and General Linear Model were used on the training data to predict the readmission risk.

*Random Forest*

Random Forests (RF) is a type of decision tree that employs modified tree learning algorithm. RF at each split of the input variables during the learning phase randomly selects a subset of features. This process is also termed called "feature bagging." and helps in determining the few highly correlated attributes that significantly influences the predictors for yielding the best-fit target output with high accuracy. Usually, for any classification problem having '$x$' features, $\sqrt{x}$ (rounded down) features are used in each split. Whereas for regression type analysis it is recommended having $x/3$ (rounded down) splits and at least node size of 5.

*Support Vector Machine*

A Support Vector Machine (SVM) is a discriminative classifier that uses a hyperplane to segregate different classes either linearly or radially. In other words, in supervised learning, the algorithm produces an optimal hyperplane which classifies output into specific categories.

For SVM Linear the prediction Equation (2 shown below) for input is a dot product between the input ($x$) and each support vector ($xi$):

$$f(x) = Bo + \sum((ai)*(x, x_i)) \quad (2)$$

Equation 2 computes the inner products of an input vector ($x$) with all support vectors in training data. The coefficients *Bo* and *ai* are assessed from the training data by the acquiring algorithm.

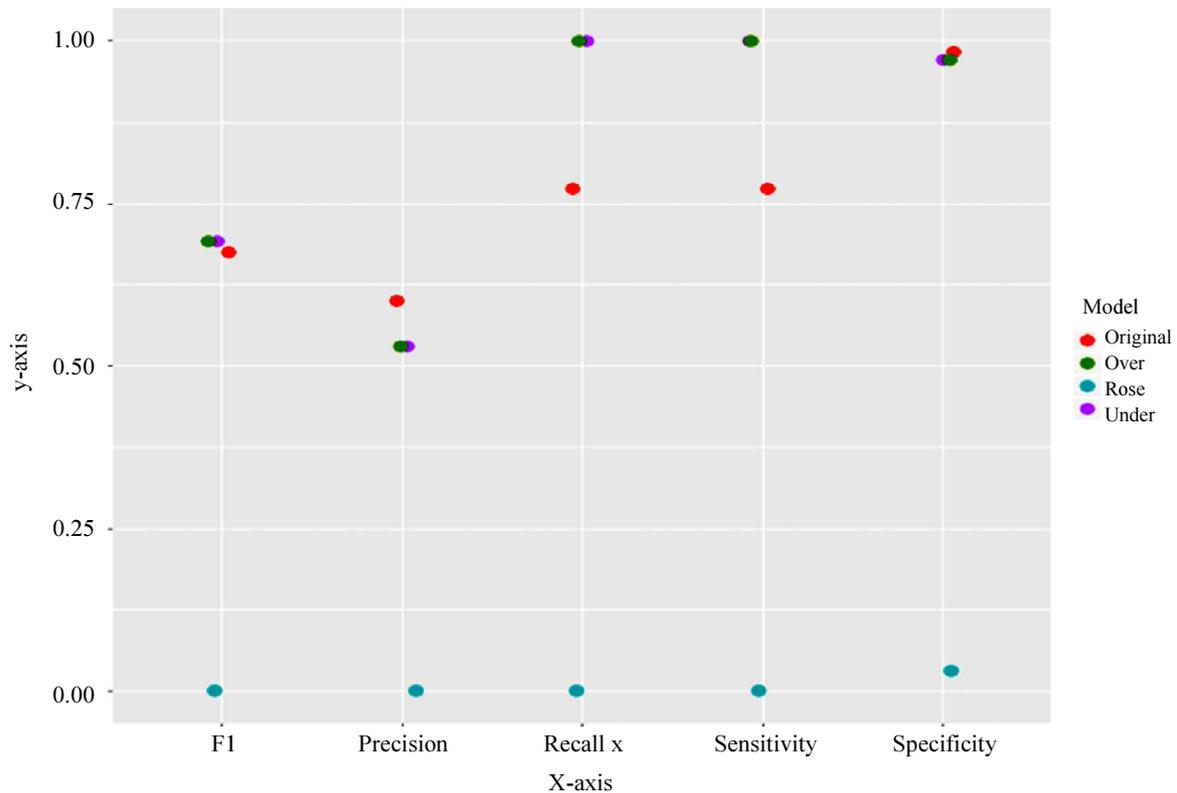

**Fig. 1:** Comparing data sampling methods. In this figure the x-axis represents the performance measure such as F1, precision, recall, sensitivity, specificity. The y-axis in the figure shows their respective value ranging from 0 and 1. Higher y value signifies better performance

SVM polynomial kernel and exponential kernel are expressed as the respective Equations (3 and 4) given below:

$$K(x, x_i) = 1 + \sum (x * x_i)^d \qquad (3)$$

$$K(x, x_i) = e^{(-\gamma * \sum((x - xi2))} \qquad (4)$$

*Gradient Boosting Method*

Gradient boosting method is a classification and regression technique that ensembles several weak prediction models and produces a decision tree. GBM like any other supervised learning defines and minimizes the loss function. The Equation (5) below shows the Loss function:

$$Loss = MSE = \sum (y_i - y^p_i)^2 \qquad (5)$$

where, $y_i = i^{th}$ target value, $y_i^p = i^{th}$ prediction, $L(y_i, y_i^p)$ is the loss function.

By implementing gradient descent, the minimum MSE can be found using the Equation (6) given below:

$$y^p_i = y^p_i - \alpha * 2 * \sum (y_i - y^p_i) \qquad (6)$$

where, is learning rate and $\Sigma(y_i - y_i^p)$ is sum of residuals.

*General Linear Model*

The general equation for the General Linear Model (GLM) is defined as the Equation (7) given below:

$$y = \beta_o + \beta_1 X_1 + \beta_2 X_2 + \ldots + \beta_{n-1} X_{n-1} + \beta_n X_n \qquad (7)$$

The $\beta_s$ in the given GLM equation are coefficients or weights dispensed to the input or predictor variables, i.e., the *X*'s on the right-hand side of the prediction equation.

*Recursive Partitioning and Regression Tree*

The Recursive Partitioning and Regression Tree (Rpart) algorithm splits the dataset recursively. In other words, the split continues till a given termination criterion is attained. It is crucial to observe that the algorithm makes the best decision at each splitting stage, without any contemplation of optimality in the upcoming stages. In other words, this approach ensures local optimality. Due to this approach, deep trees are prone to overfitting. However, overfitting can be checked by developing shallower trees by terminating the algorithm at an ideal point or by pruning the deep tree to the desired criterion. Rpart follows the later technique to minimize overfitting.

Overfitting minimization is achieved by the following Equation (8 and 9) shown below:

$$min(C\alpha(T)) \qquad (8)$$

Equation 8 minimizes the cost $C\alpha(T)$ assigned to each variable which is the linear combination (see Equation 9 below) of error $R(T)$ and the number of leaf nodes in the tree $|T|$:

$$C\alpha(T) = R(T) + \alpha|T| \qquad (9)$$

*Performance Measures*

To analyze and compare each model, we considered accuracy, sensitivity and specificity (Choudhury, 2018).

Accuracy (ACC) is determined as the number of correct prediction upon the total number of the dataset. The accuracy can vary from 0 to 1. Where 1 is the best possible result.

Sensitivity (SN) is the count of true (correct) positive predictions divided by the total number of positives. The value of SN varies from 0 to 1, where 1 is the best possible sensitivity.

Specificity (SP) is the number of true (correct) negative predictions divided by the total number of negatives. The value of SP varies from 0 to 1, where 1 is the best possible specificity.

**Table 2:** Parameters for genetic algorithm

| GA Characteristics | |
|---|---|
| Genes' alley in chromosome | Real values Chromosome length = No. of variables |
| Population | Random (uniform) population of real values Size = 15 |
| Selection Strategy | Linear-rank selection |
| Crossover Method | Local arithmetic crossover Crossover probability = 0.8 |
| Mutation Method | Uniform random mutation Mutation Probability = 0.1 |
| Replacement Strategy | Elitism by %5 |
| Termination Strategy | No. of generations = 15 |
| Constraint Handling | Constraints repair mechanism bound of [-10,10] |

*Model Optimization*

Genetic Algorithm and Greedy Ensemble algorithm were implemented to obtain the best fit model. Genetic Algorithm (GA) is a random but global search in solution space that is inspired by natural behavior of chromosomes in transmitting characteristics (genes) from one generation after another in which genes will be updated randomly with the aim of crossovers and mutations strategies through generations to produce a chromosome which is the best representative of optimal solution (Rowe, 2015). This is study considered GA for tuning a set of classifiers which provided the best performance among applied data mining methods in the previous section due to enhance the predictive accuracy and investigate the effect of classifiers' parameters on its performance measurement. Applied GA characteristics have been introduced in Table 2.

**Results**

Boruta algorithm performed 18 iterations in 1.32 h and identified 10 important attributes as shown in the Fig. 2.

However, "number of inpatients", "number of emergencies", "number of diagnoses", "diabetes med", "number of outpatients", "number of procedures" and "number of medications" were identified as the top seven influential factors by stepwise regression as shown in Fig. 3.

The important attributes identified by both Boruta and stepwise regression were (a) number of medications, (b) number of procedures, (c) number of emergencies, (d) number of outpatients, (e) number of inpatients and (f) number of diagnosis.

For further analysis we employed interaction effect study. Interaction effects study is the analysis of how multiple predictor variables, when considered together, have an impact on the main variable for analysis. It helps establish a relationship, not just between the predictors and the main variable, but also between the predictor variables themselves as shown in Fig. 4. The blue and red lines mark the low and high levels of one variable when it is being considered along with another variable. For example, in the interaction plot of the number of outpatients*number of inpatients, in a higher value setting of the number of inpatients (21), the mean of readmitted is decreasing (approaching zero) for some outpatients increases. This signifies that if every time a patient is an inpatient, he or she also becomes an outpatient, it becomes less likely that he or she will be readmitted.

Similarly, there seems to be a change in behavior when the number of inpatient and number of diagnoses is considered, wherein the number of diagnoses is the changing variable. When the number of diagnoses setting is higher, it naturally means that chances of readmission will remain high with increasing number of inpatients for a particular patient. However, even in a low setting, the mean of readmitted increases and crosses over the high setting line. This may signify that of a particular patient is an inpatient again and again, but there are no new diagnoses, the hospital may be failing concerning accuracy of judgment, which is why chances if readmission still increase.

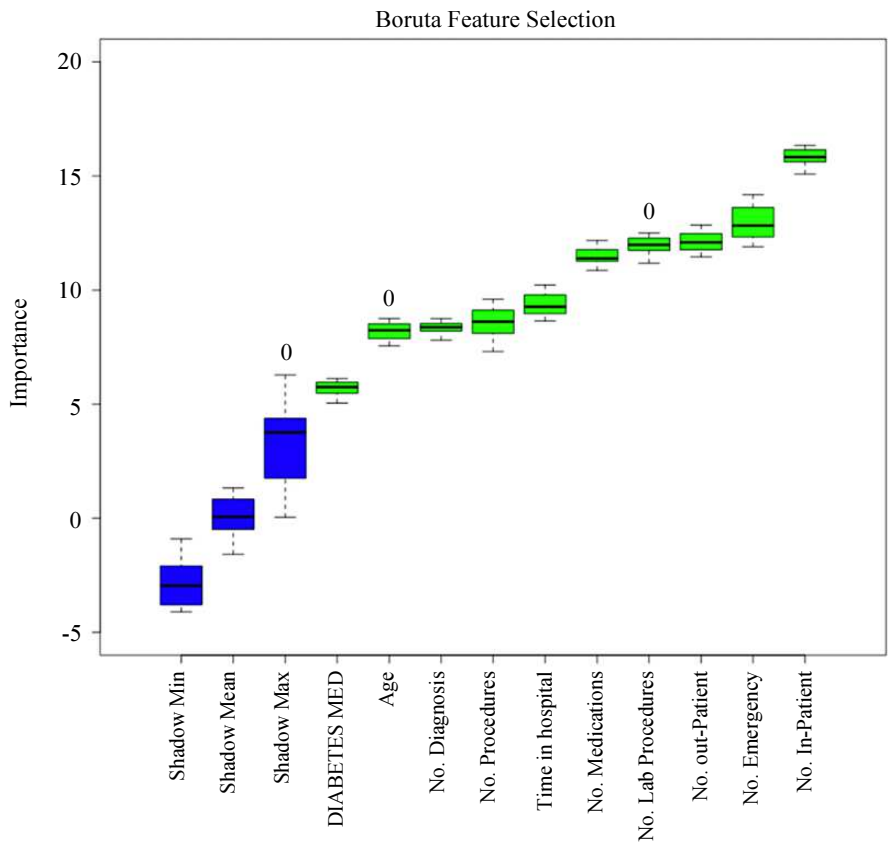

**Fig. 2:** Boruta feature selection

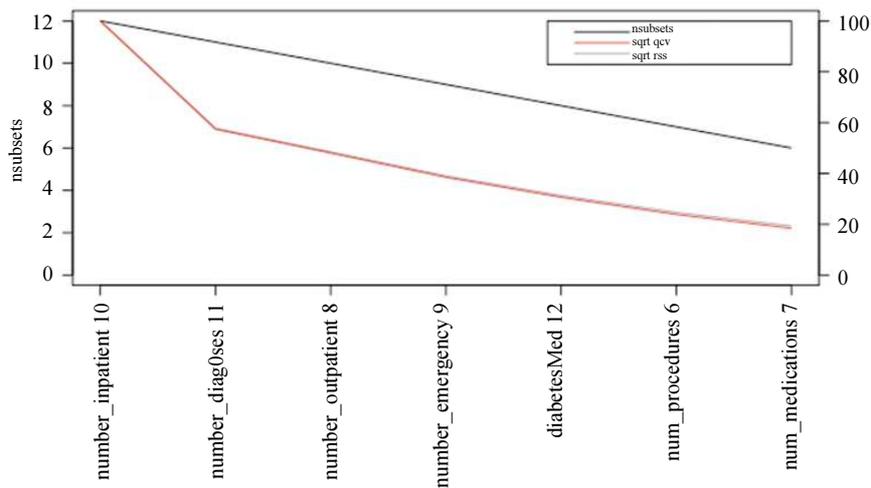

**Fig. 3**: Stepwise regression variable importance

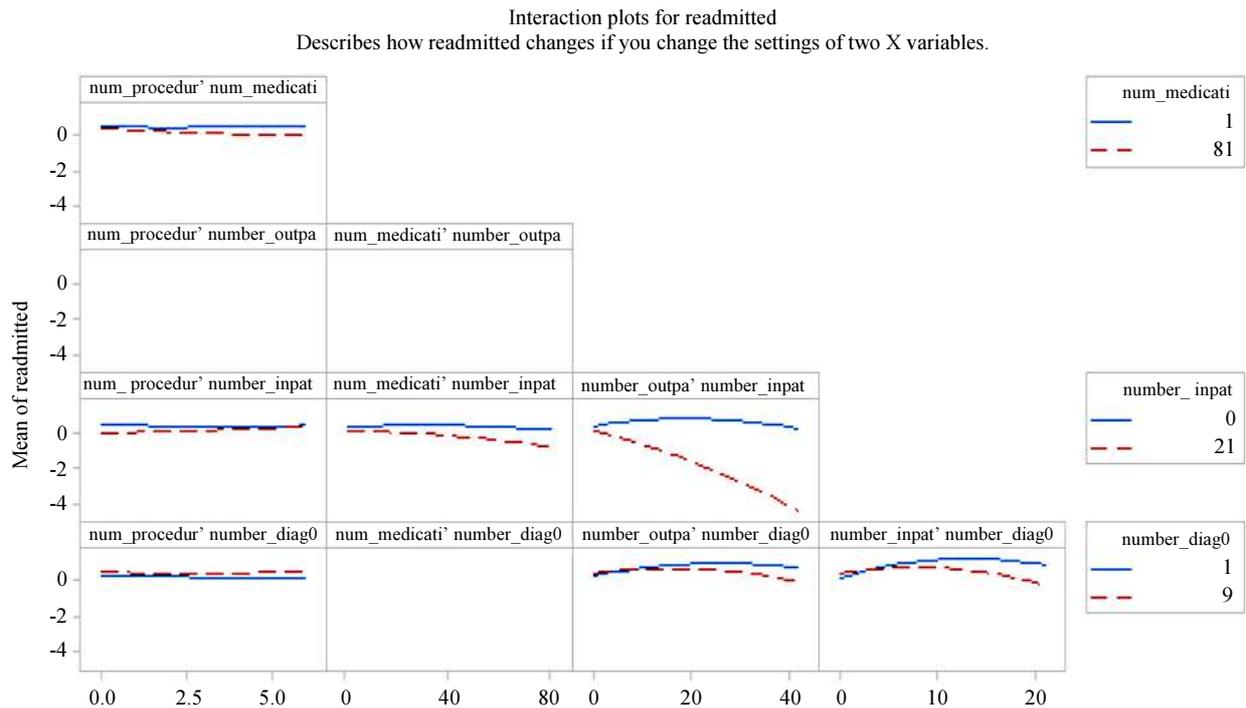

**Fig. 4:** Interaction effect analysis

**Table 3:** Model performance after data processing

| Algorithm | Accuracy (%) | Sensitivity | Specificity |
|---|---|---|---|
| Gradient boosting method | 96.12 | 0.94 | 0.97 |
| General linear model | 96.35 | 0.94 | 0.96 |
| Support vector machine | 97.00 | 0.98 | 0.96 |
| Recursive partitioning and regression trees | 96.90 | 1.00 | 0.96 |

**Table 4:** Model performance after implementing genetic algorithm

| | Gradient boosting method | General linear model | Support vector machine | Recursive partitioning and Regression Tree |
|---|---|---|---|---|
| Optimized accuracy (%) | 97.05 | 97.05 | 97.04 | 97.04 |
| Optimized classifiers' parameters | -0.69, -2.36 | 1.32, 1.64 | -4.46, 3.30 | 0.50, 5.67 |

We implemented under-sampling method to reduce the bias of the response variable. Table 3 shows the model performance and testing accuracy of all the selected algorithms.

To further enhance the performance of the models, the Genetic Algorithm was used to obtain the optimized performance as shown in the following Table 4.

Since using Genetic Optimization gave two best models with same output, we performed greedy ensemble to find the best one model.

Gradient Boosting Method was found to be the best after both Genetic Optimizations as well as Greedy Ensemble Method and we recommend GBM with 98.50% prediction accuracy as the best fit model for this dataset. Moreover, "number of inpatients" was found to be the most influencing factor that determines patient readmission risk. However, patient age, diabetes, time spent in the hospital, number of lab procedures, number of outpatients and the number of emergencies had significant relevance.

## Discussion and Conclusion

Readmission rate is a quality evaluation metric customarily used to extrapolate the quality of life index of patient population and the quality of healthcare delivery (Shameer *et al.*, 2017). Irrespective of the developments in biomedical and healthcare research practices, hospital quality control offices still use traditional predefined sets of variables to infer the probability patient readmission (Shameer *et al.*, 2017). However, predictive analytics could provide evidences to improve the quality of healthcare delivery. Uniting predictive analytics with preventive measures would involve patients, physicians and payers to contribute proactively in taming the health and wellness

In this study, we implement a predictive analytical approach to identify patients prone to readmission and thus, systematically reduce the number of avoidable readmissions mainly caused by patient non-compliances to medication instruction or early discharge from hospital. Our proposal has the capability of capturing both patient and population-based variations of hospital readmissions. It incorporates patient with diverse health concerns across 130 US hospitals. The novelty of our method is to directly incorporate patients' history of readmissions into modeling framework along with other demographic and clinical characteristics. We also verify the effectiveness of the proposed approach by validating training accuracy. Some contributions made in this paper are (i) applying Boruta algorithm and stepwise variable selection and (ii) implementing genetic and greedy ensemble algorithm to optimize our predictive models.

Our study recommended optimized gradient boosting method for identifying patient most likely to get readmitted. Furthermore, the study also emphasizes on the effectiveness of data preprocessing. It measures the influence of data balancing, removing outliers and imputing missing values on the classification accuracy. Our study also produces highest readmission prediction accuracy.

Some research directions can be sought by trying different variable selection techniques such as LASSO or Nonnegative Garrote for better subset regressions. Also, in presence of high right censored data, it is interesting to consider some health care cost measures from which it may be possible to statistically estimate the mean population cost for readmission.

## Author's Contributions

**Avishek Choudhury:** Research, data collection, analysis, data interpretation, figure formation, coding and writing manuscript.

**Dr. Christopher M. Greene:** Manuscript writing and formatting.

## Ethics

All data was collected with the permission of the organization and patient's medical and personal information was secured.

## Data Statement

The data used in this study can be retrieved from DOI: 10.17632/nntck7ddgt.1. URL: https://data.mendeley.com/datasets/nntck7ddgt/2